\documentclass[12pt]{article}
\pdfoutput=1
\usepackage{amssymb,amsmath}
\usepackage{graphicx}
\usepackage{color}
\usepackage[colorlinks=true
,urlcolor=blue
,citecolor=blue
,linkcolor=blue
,linktocpage=true
,pdfproducer=medialab
]{hyperref}
\usepackage[a4paper,width=15.3cm]{geometry}

\begin{document}
\begin{flushright} MIFPA-14-20\end{flushright}

\begin{center}

 {\Large\bf  3.5 keV X-ray line and R-Parity Conserving Supersymmetry
 } \vspace{1cm}

{\bf  Bhaskar Dutta$^a$, Ilia Gogoladze$^{b,}$\footnote{E-mail:
ilia@bartol.udel.edu\\ \hspace*{0.5cm} On  leave of absence from:
Andronikashvili Institute of Physics, 0177 Tbilisi, Georgia.},  Rizwan Khalid $^{c,}$\footnote{E-mail: rizwan.hep@gmail.com, rizwan@sns.nust.edu.pk}
  and   Qaisar Shafi$^{b,}$\footnote{E-mail:
shafi@bartol.udel.edu} } \vspace{.5cm}

{ \it
$^a$Mitchell Institute of Fundamental Physics and Astronomy, Department of Physics and Astronomy,
Texas A\&M University, College Station, TX 77843-4242, USA
$^b$Bartol Research Institute, Department of Physics and Astronomy, \\
University of Delaware, Newark, DE 19716, USA  \\
$^c$Department of Physics, School of Natural Sciences, National
University of Sciences \& Technology, H-12, Islamabad, Pakistan\\
} \vspace{.5cm}

\vspace{1.5cm}
 {\bf Abstract}\end{center}

We present some R-parity conserving supersymmetric models which can accommodate the 3.5 keV  X-ray
line reported in recent spectral studies of the Perseus galaxy cluster and the Andromeda galaxy.
Within the Minimal Supersymmetric Standard Model (MSSM) framework, the dark matter (DM) gravitino
(or the axino) with mass of around 7 keV decays into a massless neutralino (bino) and a photon with
lifetime $\sim10^{28}$ sec. The massless bino
contributes to the effective number of neutrino species
$N_{\rm eff}$ and future data will test this prediction.
In the context of NMSSM, we first consider scenarios where  the bino is massless and
the singlino mass is around 7 keV.
We also consider quasi-degenerate bino-singlino scenarios where the mass scale
of DM particles are ~$O$(GeV) or larger. In such a scenario we require the mass gap to generate the 3.5 keV line.
We comment on the possibility of a 7 keV singlino decaying via R parity violating couplings while all other
neutralinos are heavy.

\newpage

%%%%%%%%%%%%%%%%%%%%%%%%%%%%%%%%%%%%%%%%%%%%%%%%%%%%%%%%%%%%
\renewcommand{\thefootnote}{\arabic{footnote}}
\setcounter{footnote}{0}
%%%%%%%%%%%%%%%%%%%%%%%%%%%%%%%%%%%%%%%%%%%%%%%%%%%%%%%%%%%%%

\section{\label{ch:introduction}Introduction}

Two recent independent studies~\cite{Bulbul:2014sua, Boyarsky:2014jta} based on X-ray observation data show
a photon emission line at 3.5 keV energy in the spectra from Perseus galaxy cluster and the Andromeda galaxy.
This observation can be interpreted as a possible signal of dark matter (DM) decay with the
emission of a 3.5keV photon, with the DM mass ($m_{\rm DM}$) and lifetime ($\tau_{\rm DM} $) given by,
\begin{eqnarray}
&&m_{\rm DM} \simeq 7~ {\rm keV}  \nonumber \\
&&\tau_{\rm DM} \simeq 2 \times 10^{27} -  10^{28}~ {\rm sec}.
\label{eq1}
\end{eqnarray}
A variety of explanations of this line have already been proposed~\cite{all,Liew:2014gia,Bomark:2014yja,Kolda:2014ppa}.
However, there exist just a few supersymmetric scenarios which contain such a light neutral
particle. For instance, it could be an axino~\cite{Liew:2014gia}, gravitino~\cite{Bomark:2014yja} or
neutralino (bino) \cite{Kolda:2014ppa}.
These particles are able to produce
the observed X-ray line \cite{Bulbul:2014sua, Boyarsky:2014jta}  by decaying through R-parity
violating processes~\cite{Kolda:2014ppa} to a photon and  neutrino, for example.

In this paper we present some simple scenarios which can accommodate the 3.5 keV X-ray line
in the context of R-parity conserving supersymmetry (SUSY).
They include the minimal supersymmetric standard model (MSSM) and Next-to-MSSM (NMSSM).
It is interesting to note that in the MSSM, the lightest  neutralino can be
massless ~\cite{Bartl:1989ms, Gogoladze:2002xp, Dreiner:2009ic} while satisfying the current experimental constraints.
In order to realize this scenario \cite{Bartl:1989ms}, we assume that the soft
supersymmetry breaking (SSB)  MSSM  gaugino masses are arbitrary, and we impose
the requirement that the neutralino mass matrix at the weak scale has zero determinant. This can be achieved  by
suitable choice of parameters, while having very small ($\lesssim 1$ eV) or even zero mass bino,
with the  charginos (and the next to lightest neutralino $\tilde\chi^0_1$)
heavier then 420 GeV to satisfy the mass bounds on the chargino from LHC ~\cite{ATLAS}.
In our scenarios where the `near massless' bino is accompanied by a 7~keV gravitino, axino, or singlino which
behave as warm DM.
arising in different models  around keV scale giving rise to warm DM. The 7~keV DM particle decays to a
bino and a photon with an appropriate long lifetime to explain the observed X-ray line.
The warm dark matter scenario which is under investigation for a long time ~\cite{Blumenthal:1982mv},  proposes solution to the missing satelite problem
of the local group of galaxies~\cite{Bode:2000gq}. The massless bino
contributes {to the effective number of neutrino species},  $N_{\rm eff}$, which is
expected to be strongly constrained in the near future. We also consider an almost degenerate bino-singlino
scenario in the NMSSM framework, such that the mass scale of cold DM particles are ~O(GeV)
or larger.

We can retain gauge coupling unification in the presence of  non-universal gaugino masses at $M_{\rm GUT}$, which are
realized via non-singlet $F$-terms compatible with the underlying grand unified theory (GUT)~\cite{Martin:2009ad}.
Nonuniversal gauginos can also be generated from an $F$-term which is a linear combination of two
distinct fields of different dimensions \cite{Martin:2013aha}. It is also possible
to have non-universal gaugino masses \cite{Ajaib:2013kka}  in the  SO(10) GUT with unified Higgs
sector~\cite{Babu:2006rp}, or utilize two distinct sources for supersymmetry
breaking \cite{Anandakrishnan:2013cwa}. In general, in the gauge mediated supersymmetry breaking (GMSB)
scenario, all gaugino masses  can be independent of each other \cite{Meade:2008wd}.
With so many distinct possibilities available for realizing non-universal gaugino masses
while keeping universal sfermion mass ($m_{0}$) at $M_{\rm GUT}$, we employ  non-universal
masses for the  MSSM  gauginos in our study without further justification.

One of the  motivations for non-universal gauginos can be related to the
interplay between the 125 GeV Higgs boson and the explanation of the apparent muon g-2 anomaly. A universal
SSB mass term for sfermions ($m_{0}$) is needed
to suppress flavor-changing neutral
current  processes \cite{Martin:1997ns}. On the other hand, in order to accommodate the
125 GeV \cite{:2012gk, :2012gu} light CP even Higgs boson mass  and to resolve the discrepancy
between the SM and the measurement of the anomalous magnetic moment of the muon \cite{Hagiwara:2011af} in the
framework of  universal sfermion SSB masses, we need to have non-universal  gaugino masses
at $M_{\rm GUT}$  \cite{Gogoladze:2014cha}.

The outline of our paper is as follows. In section 2, we discuss the 3.5 keV line in the context of
MSSM scenarios. In section 3, we discuss possible NMSSM scenarios, followed with our conclusion in Section 4.
In the Appendix we present technical details regarding two massless neutralinos in the NMSSM and provide a few
representative solutions of interest.

\section{MSSM}

In this section, we outline several scenarios that can accommodate a 3.5 keV X-ray line in the
MSSM. Let us start by examining how it might be possible to
 obtain a massless neutralino in the framework of the MSSM.
 The neutralino mass matrix in the gauge eigenbasis
 $\Psi^0 = (\tilde B, \tilde W^0, \tilde H_d^0, \tilde H_u^0 )^T$ has  the form \cite{Martin:1997ns}
\begin{eqnarray}
\label{eq3}
\mathcal{M}_{\tilde{\chi}^0} &=&
\begin{pmatrix}
M_1 & 0   & - M_Z s_{\rm w} c_{\beta} &  M_Z c_{\rm w} s_{\beta} \\
0   & M_2 &  M_Z c_{\rm w} c_{\beta}  & -M_Z c_{\rm w} s_{\beta} \\
 - M_Z s_{\rm w} c_{\beta} & M_Z c_{\rm w} c_{\beta}  & 0 & -\mu\\
 M_Z s_{\rm w} s_{\beta} & -M_Z c_{\rm w} s_{\beta} & -\mu & 0
\end{pmatrix}\,.
\end{eqnarray}
Here $M_1, M_2 $ are the supersymmetric gaugino mass parameters
for the $U(1)$ and $SU(2)$ sector respectively, while $\mu$  is the bilinear Higgs mixing parameter.
$M_Z$  denotes the $Z$ gauge-boson mass and $s_{\rm w} \equiv\sin\theta_
{\rm{w}}$,  $c_{\rm w} \equiv\cos\theta_{\rm{w}}$, where
$\theta_ {\rm{w}}$ is the weak mixing angle.  $s_{\beta}\equiv\sin{\beta}$,  $c_{\beta}\equiv\cos{\beta}$,  while
$\tan\beta$ is the ratio of the vacuum expectation values (VEVs) of the MSSM Higgs doublets.

To realize a massless netralino~ \cite{Bartl:1989ms, Dreiner:2009ic}, the following
relation must be satisfied:
\begin{eqnarray}
M_1 = \frac{M_2 M_Z^2 \sin(2\beta)s^2_{\rm w}}{\mu M_2 - M_Z^2\sin(2\beta)c^2_{\rm w}} \approx \frac{2 M_Z^2 s^2_{\rm w}}{\mu \tan\beta}.
\label{eq5}
\end{eqnarray}
Implementing the chargino mass bound $(|\mu|, ~ M_2) > 420$ GeV  in Eq. (\ref{eq5}) leads to
$M_1\ll (M_2, \, |\mu|)$. In the Appendix we give one example of an MSSM scenario with very small
LSP neutralino (mostly bino) mass. Such a bino is consistent with
current experimental data from LEP, structure formation etc~\cite{Dreiner:2009ic}.
The LHC provides constraints on the next to lightest neutralino, chargino, and slepton
masses when the lightest neutralino is almost massless.

The relation in Eq. (\ref{eq5}) has been obtained at tree level,
but radiative corrections do not significantly modify it. Notwithstanding radiative corrections,
since $M_1$, $M_2$ and $\mu$ are free parameters, there
is no problem to ensure that the determinant in Eq. (\ref{eq3}) is zero. Thus, it is possible to
have an essentially massless neutralino by fine-tuning the parameters in the framework of the MSSM, and an
example is presented in the Appendix.

%\textbf{\large Massless Bino and $N_{\rm eff}$}.
 \textbf{The existence of a near massless bino, however, would contribute to
$\Delta N_{\rm eff}\equiv N_{\rm eff}-N_{\rm eff, SM}= 1$.}
The reason for this is that the essentially massless bino decouples from the thermal background
around the same time as the neutrinos. The decoupling temperature also depends on the slepton mass which we take around the weak scale. However, if the slepton mass increases, the decoupling temperature also increases, e.g., if the slepton mass is 10 TeV, then the decoupling temperature will be O(GeV).
The present  observational bound on $\Delta N_{\rm eff}$ from
Planck + WMAP9 + ACT + SPT + BAO + HST at 2$\sigma$ is
$\Delta N_{\rm eff} = 0.48^{+0.48}_{-0.45}$~\cite{planck}.
The value of $N_{\rm eff}$ depends on Hubble constant where there is a discrepancy between Planck and
HST~\cite{hst}. A reconciliation can occur using larger $\Delta N_{\rm eff}$~\cite{Wyman:2013lza}.
The new BICEP2 data~\cite{bicep2} also requires a larger $\Delta N_{\rm eff}$(=0.81 $\pm 0.25$) in order
to reconcile with the Planck data~\cite{Dvorkin:2014lea}. Future data hopefully will settle this issue.

\subsection{Gravitino dark matter and massless bino}

One way to  accommodate a 3.5 keV X-ray line via a massless neutralino comes from the
gauge mediated SUSY breaking (GMSB) scenario. As a consequence of the flavor blind gauge
interactions responsible for generating the SSB terms
\cite{Giudice:1998bp}, this senario provides a compelling
resolution of the SUSY flavor problem.  In both the minimal \cite{Giudice:1998bp}  and
general \cite{Meade:2008wd} GMSB versions,
the gravitino, which is the spin 3/2 superpartner of the graviton, acquires mass
through spontaneous  breaking of local supersymmetry. The gravitino mass
can be $\sim 1 {\rm \ eV} - 100 {\rm \ TeV}$. Additionally, in the general GMSB scenario,
the  SSB mass terms for the  MSSM gauginos are arbitrary. In particular, it is possible to have
a massless neutralino (essentially a bino)
in this framework. With all other sparticles being much heavier,
the gravitino dominantly decays to the neutralino (bino)
and photon ($\tilde G \rightarrow \tilde{\chi}^0_1 + \gamma$).

\begin{figure}
  \centering
  \includegraphics[width=0.3\textwidth]{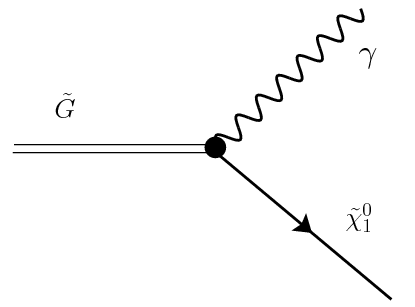}
  \caption{$\widetilde G \rightarrow \tilde\chi^0_1 + \gamma$  decay.
  \label{figure1}
}
\end{figure}

The relevant diagram for this decay is shown in Figure \ref{figure1}, and the
decay rate is given by~\cite{thesis}
\begin{eqnarray}
\label{eq23}
\Gamma (\widetilde  G \rightarrow \tilde{\chi}^0_1 \, \gamma) = \frac{\cos\theta_W^2 m^3_{\widetilde G}}{8\pi M^2_{P}}.
\end{eqnarray}
{Using Eq. (\ref{eq23}) and assuming the
gravitino mass to be 7 keV, the gravitino lifetime is estimated to
be $3 \times 10^{29}$ sec, which is approximately
a factor of 10 more than what we need which can be difficult to obtain.
%This estimate assumes, of course, that in 4D the
%reduced Planck mass is $M_{P}=2.4 \times 10^{18}$ GeV.
However, physics around the Planck scale
$M_P$ is largely unknown. 
It has been noted in ref. \cite{Dvali:2007hz,Dvali} that the fundamental
mass scale ($M_{\Lambda}$) can be reduced to $M_P/\sqrt{N}$ in the presence of a nonzero
number of degrees of freedom ($N$). In fact, it is shown that the  scale for quantum gravity in 4D becomes the new scale $M_\ast$ where the classical gravity becomes very strong and below this scale no quasi-classical black hole can exist. This becomes the scale of the non-renormalizable operators as well since this mass scale marks the new cutoff.  In this way the cutoff scale can be reduced as required in Eq. (\ref{eq23}).

It is possible to envision a larger effective coupling $\tilde G \tilde\chi^0_1 \gamma$ coupling by assuming new particles providing additional contributions to the effective $\tilde G \tilde\chi^0_1 \gamma$ coupling. For example, there could be a new operator  $\tilde G \tilde\chi^0_1 \gamma f_{\rm scalar}/M_{\Lambda}$, which can arise from the fundamental interactions, $\tilde G \gamma  f_{\rm fermion}$ and $f_{\rm scalar}\tilde\chi^0_1 f_{\rm fermion}$. By integrating the fermion $f_{\rm fermion}$ at the scale $M_{\Lambda}$ we can get the above operator. The scalar $f_{\rm scalar}$ can have a VEV $<f_{\rm scalar}>\sim M_{\Lambda}$ to give us  a new tree-level O(1) contribution to the $\tilde G \tilde\chi^0_1 \gamma$ coupling.
 It is possible to have  large contributions from many such diagrams to induce a large effective coupling to  yield the desired
lifetime for the gravitino as needed in Eq. (\ref{eq23})}. However, SUSY needs to be broken in order to  preserve  equivalence principle.

One important issue for gravitino dark matter is the reproduction of the
correct dark matter relic density. The initial thermal abundance is diluted
because of a late reheat temperature ($T_R$) arising from heavy field/moduli decay.
The relic density ($\Omega_{\tilde G} h^2$) of gravitinos which arise
from the scattering of gluinos, squarks etc.
is given by~\cite{Bolz:2000fu,Moroi:1993mb},
%%%
\begin{equation}\label{eq:ReheatTemp}
  \Omega_{\tilde G} h^2\approx0.27 \left(\frac{100 \text{ GeV}}{ m_{\widetilde G}}\right)\left(\frac{T_R}{10^{10}\text{ GeV}}\right)\left(\frac{m_{\tilde g}}{1 \text{ TeV}}\right)^2 \left(\frac{2.4\times 10^{18}\, {\rm GeV}}{ M_{\Lambda}}\right)^2.
\end{equation}

To realize $\Omega_{\tilde G} h^2\approx0.1$ with $m_{\tilde g}\gtrsim 1.4$ TeV
and $M_{\Lambda} \approx 10^{17}$ GeV, we require $T_R\lesssim 10^4$ GeV.

\subsection{Axino dark matter and massless bino}

\begin{figure}
  \centering
  \includegraphics[width=0.4\textwidth]{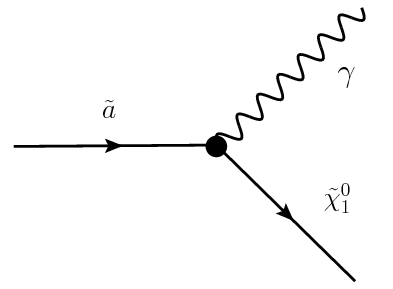}
  \caption{$\tilde a \rightarrow \tilde{\chi}^0_1 + \gamma$  decay.
  \label{figure2}
}
\end{figure}

A very compelling way of solving the strong CP problem is via the Peccei-Quinn
(PQ) mechanism \cite{Peccei:1977hh}, which yields a light pseudo-scalar field
(axion $a$) associated with the spontaneously
broken global $U(1)$ symmetry. An inevitable prediction
from a combination of PQ mechanism and low scale supersymmetry is the existence
of the supersymmetric partners of the axion, the axino ($\overline a$) and
saxion $s$ \cite{Nilles:1981py}. The axion superfield $A$ can be expressed as,

\begin{equation}
A=\frac{1}{\sqrt 2} (s+i a) + \sqrt 2\,  \tilde{a}\, \theta + F_{A} \,\theta \,\theta,
\end{equation}
where $F_A$ denotes the auxiliary field and $\theta$ is a Grassmann  coordinate.
In general, the axino mass is  very model dependent \cite{Kim:2012bb} and can lie anywhere
from eV to multi-TeV.
It was shown that a stable axino with keV mass is a viable warm dark matter
candidate \cite{Rajagopal:1990yx,Covi:2001nw}. The 3.5~keV X-ray
line  can be explained  by a decaying axino dark matter. For this purpose,
the authors in \cite{Liew:2014gia} introduce R-parity violating couplings,
with strength $\sim10^{-1}-10^{-3}$ in order to
accommodate desired axino life time.

In this paper, we propose an alternative way to explain the X-ray line using 7 keV axino dark matter.
As mentioned above, within the MSSM framework, it is possible to have a massless neutralino
in the spectrum which is consistent with all experimental constraints.
We know that the axino couples to the gauginos and gauge bosons via the anomaly induced term.
In particular, we are interested in the interaction of the  axino to  the bino ($\tilde B$) and
the hypercharge vector boson ($B$). This interaction takes the form \cite{Choi:2013lwa},
\begin{eqnarray}
\label{eq24}
i\frac{\alpha_Y C_Y}{16 \pi f_a} \gamma_5[\gamma^{\mu}, \gamma^{\nu}]\tilde{B}\, B_{\mu \nu}.
\end{eqnarray}
Here $\alpha_Y= Y^2/4 \pi$ is the hypercharge gauge coupling constant and
$C_Y$ is a model dependent coupling associated with the $U(1)_Y$ gauge anomaly interaction.
The axion decay constant is denote by $f_a$. The axino decays to a
neutralino (bino) and photon without requiring R-parity
violating interaction. The relevant diagram for this decay is shown in
Figure \ref{figure2}, and the decay rate is given by \cite{Covi:2001nw},
\begin{eqnarray}
\label{eq25}
\Gamma ( \tilde{a} \rightarrow \chi_1^0 \, \gamma) = \frac{\alpha_{em}^2 C^2_{a\chi \gamma}}{128 \pi^3}
\frac{m_{\tilde a}^3}{f_a^2},
\end{eqnarray}
where $m_{\tilde a}$ is axino mass,
$C^2_{a\chi \gamma}=(C_y/\cos\theta_W)Z_{11}$, and $Z_{11}$ denotes the bino part of the lightest neutralino.

The axino lifetime can be expressed as:
\begin{eqnarray}
\label{eq266}
\tau ( \tilde{a} \rightarrow \chi_1^0 \, \gamma) = 1.3 \times 10^{23} {\rm sec}
\left(\frac{f_a}{10^{12}\, {\rm GeV}}\right)^2 \left(\frac{7.1\, {\rm keV}}{m_{\tilde a}}\right)^3
\end{eqnarray}
From Eq. (\ref{eq266}) we see that we need to have $f_a\approx 10^{14}$ GeV is required.  On the other
hand, in order not to overproduce axion dark matter, we need to have $f_a\lesssim 10^{12}$ GeV is preferred.
One resolution  of  this is to  invoke a small  initial axion mis-alignment angle
$\theta \approx 0.1-0.01$ \cite{Fox:2004kb}, which yields the required axion dark matter abundance
while allowing $f_a\approx 10^{14}$ GeV.
An alternative solution~\cite{Dine:1982ah} is to add additional massive fields whose
late decay can inject substantial entropy into the universe at times after axion
oscillations  begin, but before BBN starts.

It is, furthermore, possible to have an axion-like particle (and associated axino)~\cite{Conlon:2006tq} in the
low scale spectrum, which may be obtained from string theory. Axino-like  particles can decay into a bino and  photon.
In this case the bound on $f_a$ can be more flexible and also the coefficient $C_y$ can be
suitably adjusted to be   $O(10^{-2})$ or so, since it is not tied  to the solution of the strong CP problem.

\section{NMSSM}

As shown in the previous section, in the MSSM it is possible to
have a massless bino, while keeping all other neutralinos heavier than 400 GeV.
In the NMSSM, the  neutralinos have a singlino component from the gauge singlet chiral superfield $S$ (with even
$Z_2$  matter parity) added to the MSSM with new terms in the superpotential:
\begin{align}
 W \supset  \mu H_u H_d + \lambda H_u H_d S - \frac{1}{3}\kappa S^3,
\end{align}
 $H_u$ and $H_d$ are the standard MSSM Higgs doublets and $\kappa$ and $\lambda$ are dimensionless couplings.
 Once the $S$ field acquires a VEV  $\langle S\rangle$, we obtain an effective $\mu$-term for
 MSSM Higgs fields, $\mu_{\rm eff}=\mu + \lambda \langle S \rangle$. The neutralino mass matrix
 in the gauge eigenstate basis  $ \Psi^0 = (\tilde B, \tilde W^0, \tilde H_d^0, \tilde H_u^0,s )^T$ has the following form:
\begin{align}
  \cal{M_{N}}=
 \begin{pmatrix}
  M_1 & 0 & -m_Z c_\beta s_W & m_Z s_\beta s_W & 0 \\
  0 & M_2 & m_Z c_\beta c_W & -m_Z s_\beta c_W & 0 \\
  -m_Z c_\beta s_W & m_Z c_\beta c_W & 0 & -\mu_{eff} & -\lambda v s_\beta \\
  m_Z s_\beta s_W & -m_Z s_\beta c_W & -\mu_{eff} & 0 & -\lambda v c_\beta \\
  0 & 0 & -\lambda v s_\beta & -\lambda v c_\beta & 2\kappa \langle S \rangle
 \end{pmatrix}.
 \label{nmssm_m}
\end{align}
It was shown in~\cite{Gogoladze:2002xp} that a massless neutralino requires that
\begin{equation}\label{k_val}
\kappa = \lambda \frac{1}{2} \left( \frac{\lambda v}{\mu}
\right)^2 \frac{0.6 m_z^2 M_2 - 0.5 \mu M_2^2 \sin2\beta}{- \mu
M_1 M_2}.\nonumber
\end{equation}
 This solution is obtained for the case when  $(\mu_{eff},\, M_1,\, M_2)> M_Z$ and the singlino is the
 lightest neutralino. We can, however, easily make the lightest neutralino to be  mostly bino
 and the next to lightest neutralino essentially the singlino.
The technical details for obtaining two massless neutralinos in the framework of NMSSM are given in appendix A.

In order to explain the 3.5 keV X-ray line,  we propose that one of the neutralinos,
which is mostly bino, is almost a massless ($\lesssim 1$  eV) particle and does  not, therefore,
contribute to the warm or cold dark matter relic abundance. The second neutralino, in this scenario,
is mostly singlino with a mass of 7 keV and gives rise to the correct dark matter relic
abundance~\cite{McDonald:2008ua}. The annihilation  of thermal NMSSM Higgs produce singlinos, and
it was shown that the correct relic abundance requires the singlino mass to be a few keV. Thus,
\begin{equation}\label{eq:singlino}
  \Omega_{\tilde \chi} h^2\approx\frac{4 (1.2)^2}{\pi^5} \left(\frac{(\kappa\lambda/3+\lambda^2) v^2 sin2\beta}{M_s M_{\tilde\chi}}\right)^2\frac{g(T_\gamma)}{g(T_R)}\left(\frac{T_RT^3_\gamma}{k_Tv^4sin^22\beta}\right)^2 \frac{M^3_{\tilde\chi}M_{pl}}{ \rho_c}.
\end{equation}
Here $M_s$ is the mass of the scalar singlet, $g(T_R)=228.75$, $g(T_\gamma)=2$, $T_R\sim 10^2-10^5$ GeV,
$k_T=(4 \pi^3 g(T)/45)^{1/2}$ and $T_{\gamma}$ is the present CMB temperature.
Choosing $\kappa=3 \times 10^{-2}$, $\lambda= 10^{-10}$, $M_1=0.23$ GeV and
$M_2=-\mu=-550$ GeV (shown in point 1 of Table 1 in the Appendix), we can have the
masses for the lightest neutralino (mostly bino) and the next to
lightest neutralino (mostly singilino) to be essentially massless and 7 keV respectively.
This scenario satisfies the dark matter relic abundance constraint.

\begin{figure}
  \centering
  \includegraphics[width=0.4\textwidth]{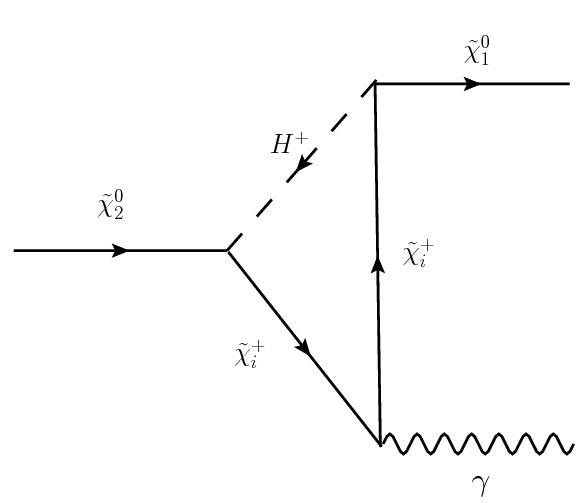}
  \hspace*{1.5cm}
  \includegraphics[width=0.4\textwidth]{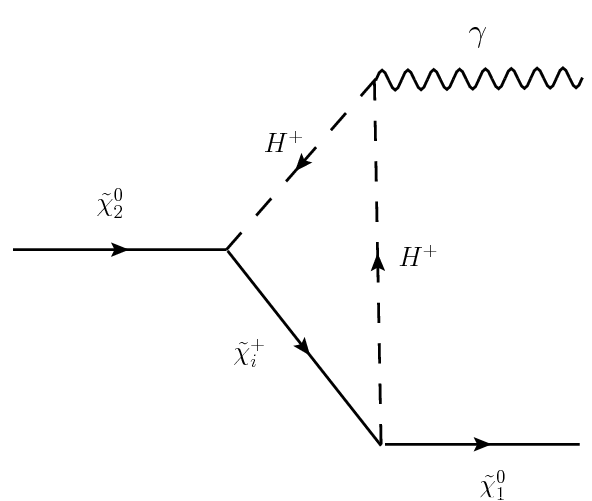}
  \caption{Decay of NLSP neutralino to the LSP neutralino with the associated emission of a photon.
  \label{figure4}}
\end{figure}

The singlino can radiatively decay to a bino and photon with a long lifetime,
which allows us to obtain the 3.5 keV X-ray line. The relevant diagram~\cite{Ellwanger:1997jj}
for this decay is shown in Figure \ref{figure4} and the decay rate is given by
\begin{eqnarray}
\label{eq27}
\Gamma ( \tilde{\chi}_2^0 \rightarrow \tilde{\chi}_1^0 \, \gamma)
 \sim \frac{\lambda^2 \alpha_{em}^2}{8\pi^3} \frac{m_{\tilde{\chi}_2}^3}{M_H^2}.
\end{eqnarray}
Here we assume that the charginos  ($m_{\tilde{\chi}^{+}_i}$) and charged  Higgs $( m_{H^+})$
have approximately the same mass.

The $\tilde{\chi}_2^0 $  lifetime can be written as:
\begin{eqnarray}
\label{eq26}
\tau ( \tilde{\chi}_2^0  \rightarrow \tilde\chi_1^0 \, \gamma) \approx 2 \times 10^{27} {\rm sec}
\left(\frac{M_H}{10^5\, {\rm GeV}} \right)^2  \left(\frac{10^{-10}}{\lambda}\right)^2.
\end{eqnarray}

In the NMSSM, an alternative explanation for the $3.5\,{\rm keV}$ emission line requires one to
have two quasi-degenerate neutralinos (bino and singlino), with mass difference
arranged to be  $\sim 3.5$~keV. We present one such example in the Appendix.
We require the next to lightest supersymmetric
particle (NLSP), which is a mixture of singlino and bino, to be long-lived
on cosmological time scales. The decay of this NLSP to the LSP,
which again  may be a bino-singlino mixture, can explain the $3.5\,{\rm keV}$ emission line.

The relevant Feynman diagrams for the NLSP neutralino decay are given in Figure~\ref{figure4}.
The decay width is given by~\cite{LS},

\begin{align}
 \Gamma \sim \frac{\alpha^2_{em} \lambda^2}{64\pi^4} \frac{(\Delta m_{\chi})^3}{m^4_{H^{\pm}}} m^2_{\chi},
\end{align}
where $\alpha_{em}$ is the electromagnetic coupling constant, $m_{\chi}$ is the quasi-degenerate mass of the
two lightest neutralinos, $\Delta m_\chi$ is their mass splitting, and
$m_{H^{\pm}}$ is the mass of the charged Higgs.

%%%%%%%%%%%%%%%%%%%%%%%%%%%%%%%%%%%%%%%%%%%%%%%%%%%%%%%%%

Assuming $m_{\chi^0_2}\approx m_{\chi^0_1}\approx 1$ GeV and $\Delta m_{\chi}\approx 3.5$ keV,
and as an example we consider   $\lambda\approx 10^{-8}$   and $m_{H^{\pm}}=500$ GeV in order to have
$\tau ( \chi^0_2 \rightarrow \chi_1^0 \, \gamma)\approx (10^{27} - 10^{28})$ sec.
The dark matter in this case is cold compared to the previous scenarios.

%The magnitude of $\Lambda$ depends on the relative phase between %$m_{\chi^0_1}$ and $m_{\chi^0_2}$. When they have opposite %phases, $\Gamma \sim \frac{\alpha^2_{em} \lambda^2}{64\pi^4} %\frac{(\Delta m_{\chi})^5}{m^4_{H^{\pm}}}$. In this case  %$\lambda\approx 10^{-3}$   is needed for $m_{H^{\pm}}=500$ GeV in %order to have
%$\tau ( \chi^0_2 \rightarrow \chi_1^0 \, \gamma)\approx (10^{27} - %10^{28})$ sec.

The singlino/bino dark matter can be produced non-thermally
from the decay of some heavy field/moduli ($\phi$) with a reheat temperature $\gtrsim 2$ MeV
in order to avoid problems with big bang nucleosynthesis. As shown in~\cite{rouzbeh},
if the abundance of DM production (combination of  dilution factor due to decay and branching
ratio into DM particles) is small enough to satisfy the
DM content, the annihilation cross-section of dark matter becomes irrelevant.

The DM abundance is given as
$n_{\rm DM}/s=min[(n_{\rm DM}/s)_{\rm obs}(3\times 10^{26}/<\sigma v>_f)(T_f/T_R), Y_{\phi}\, Br_{\rm DM}]$,
where $(n_{\rm DM}/s)_{\rm obs}\simeq 5\times 10^{-10} (1 \,{\rm GeV}/m_{\rm DM})$, $T_R$ is
the reheat temperature, $Y_{\phi}=3 T_R/4 m_{\phi}\simeq 1/\pi\sqrt{c  m_\phi/M_P}$, and $BR_{\rm DM}$
denotes the branching ratio for $\phi$ decay into singlino/bino. The singlino DM does not reach thermal
equilibrium after production from the decay of the heavy field since the decoupling temperature is
much larger than the reheat temperature $T_R$.

It is also interesting to note that the singlino can be the
lightest sparticle, and it can then decay via some R-parity violating couplings.
We present an example in the Appendix.
A slight change in the parameter values corresponding to the existence of massless neutralinos
will make the neutralino mass around keV.
A keV scale singlino LSP can decay at loop level in the presence of R-parity violating couplings.
Here we consider only the lepton number violation operators:
\begin{eqnarray}
{\cal
L}_{\not{R}}=\lambda_i L_i H_u S + \lambda_{ijk} L_i L_j E^c_k+ \lambda^{\prime}_{ijk}Q_i L_j d^c_k + \mu_i H_u L_i.
\end{eqnarray}
 The neutralino-neutrino mass matrix in the gauge eigenstate
 basis  $\Psi^{0^T} \equiv (\tilde B^0, \tilde W^0_3, \tilde h^0_d, \tilde h^0_u, \tilde s, \nu_i)$ is given by
\begin{eqnarray}
{\cal M}_{\tilde
\chi^0} = \left(\begin{array}{cc}
{\cal M_N}  & \xi_{\not{R}}^{T} \\
\xi_{\not{R}} & {\cal M}_{3 \times 3}^{\nu}
\end{array} \right),
\end{eqnarray}
where
\begin{eqnarray}
\xi_{\not{R}} = \left(\begin{array}{ccccc}
-\frac{g'v_1}{\sqrt{2}} & \frac{gv_1}{\sqrt{2}} & 0 & \mu_1 +\lambda_1 \langle s \rangle & \lambda_1 v_u  \\
-\frac{g'v_2}{\sqrt{2}} & \frac{gv_2}{\sqrt{2}} & 0 & \mu_2 +\lambda_2 \langle s \rangle & \lambda_2 v_u  \\
-\frac{g'v_3}{\sqrt{2}} & \frac{gv_3}{\sqrt{2}} & 0 &
\mu_3+\lambda_3 \langle s \rangle & \lambda_3 v_u
\end{array} \right),
\end{eqnarray}
and ${\cal M}_{3 \times 3}^{\nu}$ is the $3\times3$ light neutrino majorana mass matrix.

\begin{figure}
  \centering
  \includegraphics[width=0.4\textwidth]{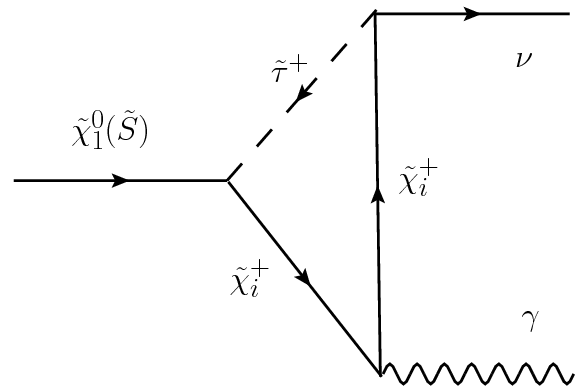}
  \hspace*{1.5cm}
  \caption{Decay of LSP neutralino through R-parity violation term.
\label{figure3}}
\end{figure}

One  of the dominant  diagrams for the decay
$\tilde{\chi}^0_1 \rightarrow \nu + \gamma$ is  given in Figure~\ref{figure3}, and the corresponding
decay rate is given by
\begin{eqnarray}
\Gamma ( \tilde{\chi}_1^0 \rightarrow \nu \, \gamma)
 \sim \alpha_{em}\frac{(\lambda \lambda_1 )^2 }{32\pi^3} \frac{\tilde{\chi}_1^3}{M_H^2}.
\end{eqnarray}
Here we assume, for simplicity, that the charged Higgs and charginos have similar masses
$M_H\equiv (m_{\tilde{\chi}^{+}_i}\approx m_{H^+})$. The singlino lifetime can be expressed as
\begin{eqnarray}
\tau ( \tilde{\chi}_1^0  \rightarrow \nu \, \gamma) \approx 2 \times 10^{27} {\rm sec}
\left(\frac{M_H}{10^5{\rm GeV}} \right)^2  \left(\frac{10^{-11}}{\lambda_1 \lambda}\right)^2,
\end{eqnarray}
and if we assume $\lambda_1\approx \lambda \approx 3 \times 10^{-6}$, the desired singlino life time is
obtained. The LSP singlino, as mentioned above, can provide the correct DM abundance.

\section{Conclusion}
In summary, we have presented several scenarios that can accommodate the
3.5 keV X-ray line in the context of R-parity conserving SUSY.
In the MSSM, the LSP neutralino can be massless and the gravitino
or axino dark matter of mass around 7 keV can decay into the LSP neutralino and a
photon with lifetime $\sim 10^{28}$ sec.
To realize this scenario, we assume that the soft SUSY
breaking MSSM gaugino masses are non-universal and they satisfy the requirement that the
determinant of the neutralino
mass matrix vanishes at the weak scale. This can always be achieved with a suitable choice of parameters,
while keeping the  charginos
(and second lightest neutralino $\tilde\chi^0_2$) heavier than 420 GeV to avoid the LHC constraint.
A keV mass dark matter is of considerable interest since it can provide
potential solutions to the missing satellites problems of the Local Group of Galaxies. The massless bino, however,
contributes to $N_{\rm eff}$ and
future data should seriously test this scenario. In the context of NMSSM, we
consider scenarios where the bino is massless and the dark matter singlino mass is around 7 keV.
Within the NMSSM, we also consider quasi-degenerate
bino-singlino scenarios where the DM mass scale is $O$(GeV) or larger. We require, in
this scenario, a small mass gap to generate the 3.5 keV X-ray line.
In passing, we also consider scenarios where the singlino is the lightest SUSY particle, and it
decays via R parity violating couplings which give rise to the 3.5 keV X-ray line.

\section*{Acknowledgments}

We would like to thank R. Allahverdi, Y. Gao  for very  useful discussions.
This work is supported in part by the DOE Grants Nos. DE-FG02-13ER42020 (B.D.) and DE-FG02-12ER41808 (I.G. and Q.S.).
I.G. acknowledges support from the  Rustaveli National Science Foundation  No.  31/98.

\appendix

\section*{Appendix}

\section{Two massless neutralinos in the NMSSM}

The neutralino mass matrix is given in Eq. (\ref{nmssm_m}) and we
seek a solution with two massless neutralinos. Assuming that $\gamma$ is
an eigenvalue of ${\cal M_N}$, we can write the characteristic equation in the form
\begin{align}
 \left\vert {\cal{M_N}}-\gamma I_5\right\vert = \gamma^5 + A\gamma^4 + B\gamma^3 + C\gamma^2 + D\gamma + E = 0,
 \label{eq20}
\end{align}
where $I_5$ is the $5\times5$ identity
matrix, and $A,B,C,D,E$, of course, depend on the entries in ${\cal M_N}$. It is known that
$A, B, C, D$ and $E$ are invariants (under similarity transformations) of the matrix
and, in particular, $E$ is the determinant of ${\cal{M_N}}$.
We can express the coefficients in Eq. (\ref{eq20}) in terms of the mass eigenstates:
\begin{eqnarray}
E=m_1^2m_2^2m_3^2m_4^2m_5^2; ~~~~~
D=\displaystyle\sum\limits_{i\neq j \neq k \neq l}^n m_i^2\, m_j^2\,m_k^2\,m_l^2;~~~~~
C=\displaystyle\sum\limits_{i\neq j \neq k }^n m_i^2\, m_j^2\,m_k^2;\nonumber \\
B=\displaystyle\sum\limits_{i\neq j \neq k }^n m_i^2\, m_j^2~~~~~
A=\displaystyle\sum\limits_{i=1 }^5 m_i^2.
\end{eqnarray}
A necessary and sufficient condition for any one eigenvalue to be zero is for the
determinant of the matrix to be zero ({\it i.e.} $E=0$). The quintic characteristic
equation then reduces to a quadratic one. Proceeding in this fashion,
if we now also set $D=0$, we will ensure that two eigenvalues of the mass matrix are
zero. It is then possible to adjust the parameters to get the desired small mass eigenvalues.

While the general expression for the the determinant and
the coefficient of $\gamma$ in the characteristic equation (variously
known as the fourth invariant) is rather complicated, the conditions to
obtain two massless neutralinos simplifies in the limit of large $\tan\beta$.
Setting $s_\beta\to 1$ and $c_\beta \to 0$ in the neutralino mass matrix, we
obtain the following conditions for two massless neutralinos,
\begin{align}
 D =& -M_1 M_2 (\lambda^2 v^2 +\mu^2) - 2 \kappa x\mu^2 (M_1+ M_2) + \nonumber\\
    & - 2 \kappa x m_Z^2 (M_1 c_W^2 + M_2 s_W^2) + m_Z^2 v^2 \lambda^2 =0 \nonumber \\
 E =& 2 M_1 M_2 \kappa x \mu^2 -m_Z^2 \lambda^2 (M_1 c_W^2 + M_2 s_W^2) = 0
 \end{align}
There can, however, be issues while using this approximation because of the large
differences in orders of magnitudes of the various terms.
In practice it is much simpler to numerically fine-tune the parameters in the
exact expressions to obtain two zero
eigenvalues.
We are essentially interested in a quasi-degenerate ($\lesssim 1$~GeV)
bino-singlino mixture. With $\lambda$ small, there is very little mixing between the
singlino and the higgsinos, particularly for $\mu\gtrsim 100$~GeV (which is needed
as previously explained). Furthermore, if we choose $M_1, 2\kappa x \sim 1$~GeV and
$M_2\gtrsim 400$~GeV, we should naively expect to get the required neutralino
masses.

In Table 1 we display three representative solutions that correspond to the three scenarios
for obtaining the 3.5 keV X-ray line within the NMSSM framework. Point 1 corresponds to
a massless bino with a 7 keV singlino. Point 2 shows the quasi-degenerate scenario involving the
bino and singlino, with a mass of 1GeV and a mass splitting of 3.5keV. Point 3 describes
the scenario in which the singlino is $\sim 7$ keV and all other neutralinos are heavy.

\begin{table}
\centering
\begin{tabular}{lcccc}
  \hline
  \hline
	                       & Point 1                        & Point 2                       & Point 3       \\
\hline
  $M_2$ (GeV)                  & 550                            & 500                           & 550                  \\
  $\mu$   (GeV)                & 550                            & 500                           & 550                       \\
  $x$  (GeV)                   & 0.0001                         & 1                             & $7\times10^{-6}$                    \\
  $\tan\beta$                  & 30                             & 30                            & 50                    \\
  $M_1$                        & 0.234                          & 1.267                         & 550                    \\
  $\kappa$                     &$3.5\times 10^{-2}$             & 0.5                           & 0.5            \\
  $\lambda$                    &$10^{-10} $                     & $10^{-9} $                    & $10^{-5} $   \\
\hline
  $m_{\tilde\chi_1^0}$ (GeV)   & $6.69 \times 10^{-13}$         & 1                             & $7\times10^{-6}$                 \\
  $\tilde\chi_1^0$ composition & $\simeq 100\%\tilde B$         & 99\%$\tilde B$                & $\simeq100\%$ $\tilde S$          \\
  $m_{\tilde\chi_2^0}$ (GeV)   & $7\times10^{-6}$               & 1         & 1.08                  \\
  $\tilde\chi_2^0$ composition & $\simeq100\%\tilde S$          & 99\%$\tilde S$ & mixture        \\
 \hline
   $m_{\tilde\chi_3^0}$ (GeV)  & 498                            & 445          & 550                  \\
  $m_{\tilde\chi_4^0}$ (GeV)   & 554                            & 505          & 554                    \\
  $m_{\tilde\chi_5^0}$ (GeV)   & 605                            & 560          & 617                     \\

  \end{tabular}
 \caption{Three representative solutions.  \label{table1}}
\end{table}

As far as the MSSM case is concerned, things are even simpler. For example, one could take,
$\tan\beta=30$, $M_2=\mu=550$~GeV, $M_1=0.23$~GeV where, $M_1$ is chosen to obtain a massless
bino. The masses of the three heavier neutralinos are
499 GeV, 555~GeV and 606 GeV.

\end{document}